\documentclass[12pt]{article}
\usepackage{amsmath}
\usepackage{amsfonts}
\usepackage{pstricks}
\usepackage{cite}
\usepackage[margin=1in,a4paper,includeheadfoot]{geometry}
\usepackage[small]{titlesec}
\usepackage{eucal}

\numberwithin{equation}{section}
\newcommand{\rme}{\textrm{e}}
\newcommand{\rmi}{\textrm{i}}
\newcommand{\eps}{\varepsilon}
\newcommand{\wt}{\widetilde}
\newcommand\ot{\mathop{\otimes}}

\newcommand{\bra}[1]{\langle\, #1\, \rvert}
\newcommand{\ket}[1]{\lvert\, #1\, \rangle}

\newcommand{\rtol}[1]{\substack{\displaystyle\longleftarrow\\ #1}}

\titleformat{\section}{\bf\large}{\thetitle.\ \;}{0pt}{}
\titleformat{\subsection}{\bf}{\thetitle.\ \;}{0pt}{}

\begin{document}
\begin{titlepage}
\begin{flushright}
DFF-422/01/05
\end{flushright}
\bigskip

\begin{center}
\Large\textbf{
On two-point boundary correlations
in the six-vertex model with DWBC
}
\end{center}
\bigskip

\begin{center}
\large{
F. Colomo and A. G. Pronko\footnote{On leave of absence from
St-Petersburg Department of V. A. Steklov
Mathematical Institute
of Russian Academy of Sciences,
Fontanka 27, 191023 St-Petersburg, Russia.}
}
\end{center}

\begin{center}
\textsl{
I.N.F.N., Sezione di Firenze,
and Dipartimento di Fisica, Universit\`a di Firenze,\\
Via G. Sansone 1, 50019 Sesto Fiorentino (FI), Italy
}
\end{center}

\bigskip
\centerline{\textbf{Abstract}}
\smallskip

The six-vertex model with domain wall boundary  conditions (DWBC)
on an $N \times N$ square lattice is considered.
The two-point correlation function describing the probability
of having two vertices in a given state at opposite (top and bottom)
boundaries of the lattice is calculated.
It is shown that this two-point boundary correlator is
expressible in a very simple way in terms of the one-point boundary
correlators of the model on $N \times N$ and $(N-1) \times (N-1)$
lattices. In alternating sign matrix (ASM) language
this result implies that the doubly refined $x$-enumerations of ASMs
are just appropriate combinations of the singly refined ones.

\end{titlepage}
\setcounter{page}{2}
\section{Introduction}

The six-vertex model, introduced in \cite{S-41}, and solved
for periodic boundary conditions in \cite{L-67,L-67a,S-67},
see \cite{LW-72,B-82} for a review,
has turned out to be of great interest also in the
case where domain wall boundary conditions (DWBC) are imposed.
These boundary conditions were originally introduced for the six-vertex
model in the investigation  of the norms of Bethe states \cite{K-82},
in the context of the quantum inverse scattering method (QISM)
\cite{KBI-93}.
An important result was obtained in \cite{I-87} where an exact
determinant formula for the partition function was obtained, see also
\cite{ICK-92}. Subsequently,
this result was found of fundamental importance in the proof of
long-standing conjectures in enumerative combinatorics,
due to the close connection of the model with
alternating sign matrices (ASMs)
\cite{MRR-83,EKLP-92a,EKLP-92b,Z-96a,Ku-96,Z-96b}, see also
\cite{Br-99} for a review.
It should be mentioned that ASM
enumerations  appear to be in turn deeply related with quantum
spin chains and some loop models, via Razumov-Stroganov
conjecture  \cite{RS-01}; for recent results, see for instance
\cite{dGR-04,DfZj-04,dGN-05} and references therein.

An important information is also contained in correlation functions.
However, because of difficulties caused by the lack of translational
invariance, their computation is still an open problem in the
six-vertex model with DWBC. Some
simplifications take place when correlations are considered in
vicinity of the boundaries \cite{BKZ-02}.
The simplest one-point boundary correlation
functions were investigated in \cite{BPZ-02}, where determinant
representations were obtained, analogous to that of papers
\cite{I-87,ICK-92} for the partition function.
Even if these are almost the simplest correlations one can study for the
considered model, they are nevertheless of
interest, especially from a combinatorial point of view.

In the present paper
we pursue the
investigation of the boundary correlation functions.
Here we evaluate the two-point boundary correlation
function which gives the probability of having particular vertex states
at two specific sites in the first and last row of the lattice.
Such a correlation function
was recently discussed  in \cite{FP-04},
in the context of a  graphical interpretation of the approach of
\cite{BPZ-02}.
{}From a combinatorial
point of view this correlation
function is closely related with the
doubly refined $x$-enumerations of ASMs; the case of
doubly refined $1$-enumeration of ASMs was studied previously in
\cite{S-02} where an explicit expression
for the corresponding generating function was derived.
An interest in the doubly refined enumerations,
in connection with the Razumov-Stroganov conjecture,
was recently stressed in \cite{Df-04}.
Here we give
an explicit, general, and relatively simple expression
for these quantities.

More specifically, we show that the mentioned
two-point boundary correlation function, for generic values
of the six-vertex model weights,
is expressible in a very simple way in terms of the analogous
one-point boundary
correlation functions. This implies that by specializing the
parameters of the model to the values
corresponding to weighted enumerations of ASMs, one can directly
obtain explicit formulae for the doubly refined weighted enumerations
of ASMs from our results here. In particular, the doubly refined
$1$-, $2$-, and $3$-enumerations of ASMs can be easily found
from the corresponding singly refined ones.

To derive the result we consider first the two-point
boundary  correlation function in the more general case of the
inhomogeneous six-vertex model. In this case QISM can be applied.
The homogeneous limit is performed next,
and a determinant representation, analogous to those given in
\cite{I-87,ICK-92,BPZ-02}, is derived.
Then, using standard techniques of the theory of orthogonal
polynomials, along the lines of our recent paper
\cite{CP-05}, we obtain the final formula for the two-point
boundary correlation function.

\section{The six-vertex model with DWBC and QISM}\label{sec-qism}

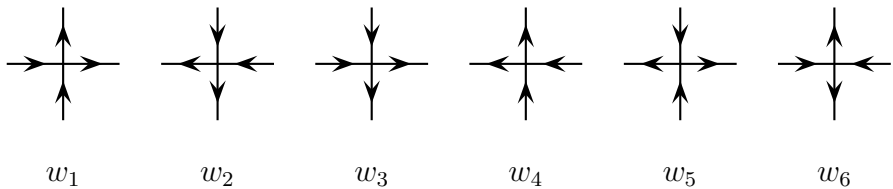
\begin{figure}[ht]
\begin{center}
\psset{unit=.75cm,arrowscale=2}
\begin{pspicture}(0,-1)(2,2)
\psline(1,0)(1,2)\psline{->}(1,.6)(1,.7)\psline{->}(1,1.6)(1,1.7)
\psline(0,1)(2,1)\psline{->}(.6,1)(.7,1)\psline{->}(1.6,1)(1.7,1)
\rput(1,-1){$w_1$}
\end{pspicture}
\quad
\begin{pspicture}(0,-1)(2,2)
\psline(1,0)(1,2)\psline{<-}(1,.3)(1,.4)\psline{<-}(1,1.3)(1,1.4)
\psline(0,1)(2,1)\psline{<-}(.3,1)(.4,1)\psline{<-}(1.3,1)(1.4,1)
\rput(1,-1){$w_2$}
\end{pspicture}
\quad
\begin{pspicture}(0,-1)(2,2)
\psline(1,0)(1,2)\psline{<-}(1,.3)(1,.4)\psline{<-}(1,1.3)(1,1.4)
\psline(0,1)(2,1)\psline{->}(.6,1)(.7,1)\psline{->}(1.6,1)(1.7,1)
\rput(1,-1){$w_3$}
\end{pspicture}
\quad
\begin{pspicture}(0,-1)(2,2)
\psline(1,0)(1,2)\psline{->}(1,.6)(1,.7)\psline{->}(1,1.6)(1,1.7)
\psline(0,1)(2,1)\psline{<-}(.3,1)(.4,1)\psline{<-}(1.3,1)(1.4,1)
\rput(1,-1){$w_4$}
\end{pspicture}
\quad
\begin{pspicture}(0,-1)(2,2)
\psline(1,0)(1,2)\psline{->}(1,.6)(1,.7)\psline{<-}(1,1.3)(1,1.4)
\psline(0,1)(2,1)\psline{<-}(.3,1)(.4,1)\psline{->}(1.6,1)(1.7,1)
\rput(1,-1){$w_5$}
\end{pspicture}
\quad
\begin{pspicture}(0,-1)(2,2)
\psline(1,0)(1,2)\psline{<-}(1,.3)(1,.4)\psline{->}(1,1.6)(1,1.7)
\psline(0,1)(2,1)\psline{->}(.6,1)(.7,1)\psline{<-}(1.3,1)(1.4,1)
\rput(1,-1){$w_6$}
\end{pspicture}
\end{center}
\caption{The six states and their Boltzmann weights.}
\label{fig-vert}
\end{figure}

In this paper we consider the six-vertex model
on an $N\times N$ square lattice with the domain wall boundary
conditions (DWBC).
Recall that the six-vertex model is a model
of arrows residing on the edges of the lattice, with the same number of
incoming and outgoing arrows though each lattice vertex (this constraint
being known as the `ice rule').
Each vertex can be in one of six possible states $i=1,\dots,6$, see
figure \ref{fig-vert}.
A Boltzmann weight $w_i$ is assigned to each vertex according to its
state $i$; the weights are usually chosen to obey the
arrow-reversal symmetry
\begin{equation}
w_1=w_2=a,\qquad
w_3=w_4=b,\qquad
w_5=w_6=c.
\end{equation}
The vertex weights are parameterized in the standard way in terms  of
a spectral parameter $\lambda$ and a crossing parameter $\eta$,
\begin{equation}\label{abc}
a=\sin(\lambda+\eta),\qquad
b=\sin(\lambda-\eta),\qquad
c=\sin(2\eta).
\end{equation}

In the case of the $N\times N$ lattice, the ice rule allows imposing
domain wall boundary conditions  to the six-vertex
model. This means fixing the direction of the boundary arrows as follows:
all arrows on the left and right boundaries
are outgoing while on the top and bottom boundaries
they are incoming.
The partition function of the model, denoted as
$Z_N$, is the sum over all
possible arrow configurations
\begin{equation}
Z_N=\sum a^{n_1+n_2} b^{n_3+n_4} c^{n_5+n_6}
\end{equation}
where $n_i$ denotes the number of vertices of type $i$ in
a configuration and $n_i$'s satisfy $n_1+\dots+n_6=N^2$.

\begin{figure}[t]
\begin{center}
\psset{unit=.75cm,dotsep=.1,arrowscale=2}
\begin{pspicture}(0,0)(7,7)
\multips(1,0)(1,0){5}{\psline(0,0)(0,2.2)}
\multips(0,1)(0,1){5}{\psline(0,0)(2.2,0)}
\multips(1,6)(1,0){5}{\psline(0,0)(0,-3.2)}
\multips(6,1)(0,1){5}{\psline(0,0)(-3.2,0)}
\multips(1,2.3)(1,0){5}{\psline[linestyle=dotted](0,0)(0,.4)}
\multips(2.3,1)(0,1){5}{\psline[linestyle=dotted](0,0)(.4,0)}
\multirput(1,.2)(1,0){5}{\psline{->}(0,0)(0,.5)}
\multirput(.2,1)(0,1){5}{\psline{<-}(0,0)(.5,0)}
\multirput(1,5.8)(1,0){5}{\psline{->}(0,0)(0,-.5)}
\multirput(5.8,1)(0,1){5}{\psline{<-}(0,0)(-.5,0)}
\rput(1,7){$\lambda_N$}
\rput(2.6,7){$\dots$}
\rput(4,7){$\lambda_2$}
\rput(5,7){$\lambda_1$}
\rput(7,1){$\nu_N$}
\rput(7,2.6){$\vdots$}
\rput(7,4){$\nu_2$}
\rput(7,5){$\nu_1$}
\end{pspicture}
\end{center}
\caption{A lattice with DWBC.}
\label{fig-dwbc}
\end{figure}
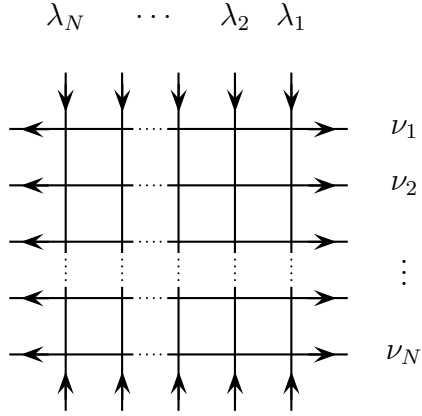

To apply the quantum inverse scattering method (QISM)
we shall consider the six-vertex model with DWBC in
its inhomogeneous version, namely, when the weights of the vertex being
at the intersection of $k$-th horizontal line (row)
and $\alpha$-th vertical line (column) are parameterized as
\begin{equation}
a(\lambda_\alpha,\nu_k)=\sin(\lambda_\alpha-\nu_k+\eta)\qquad
b(\lambda_\alpha,\nu_k)=\sin(\lambda_\alpha-\nu_k-\eta)\qquad
c(\lambda_\alpha,\nu_k)=c=\sin(2\eta).
\end{equation}
The spectral parameters
$\lambda_1,\dots,\lambda_N$ and
$\nu_1,\dots,\nu_N$ are assumed to be different
within each set.  A lattice with DWBC, and the
assignment of the spectral parameters to rows and columns
are shown in figure \ref{fig-dwbc}.
Note, that we enumerate columns (labelled by Greek indices)
from right to left and rows
(labelled by Latin indices) from top to bottom.

Apparently, the partition function (and correlation functions)
of the inhomogeneous model
is a function of $2N$ variables,
$Z_N=Z_N(\{\lambda_\alpha\},\{\nu_k\})$.
After applying QISM the spectral parameters will be set
equal within each set: $\lambda_\alpha=\lambda$ and
$\nu_k=0$.
We shall refer to this procedure as the homogeneous limit.

We shall now introduce the main objects of QISM, such as $L$-operator
and monodromy matrix.
The $L$-operator of the six-vertex model
is nothing but a matrix of the Boltzmann weights.
To each vertex being intersection of $\alpha$-th column
and $k$-th row one can associate the operator
$L_{\alpha k}(\lambda_\alpha,\nu_k)$ which acts
in the direct product of two vector spaces
$\mathbb{C}^2$: in
the `horizontal' space $\mathcal{H}_k=\mathbb{C}^2$
(associated with the $k$-th row) and in the `vertical'
space $\mathcal{V}_\alpha=\mathbb{C}^2$
(associated with the $\alpha$-th column).
The arrow states on
the top and right edges of the vertex can be viewed as
`in' indices of the
$L$-operator while those on the bottom and left edges
as `out' ones. Using spin up and spin down states
\begin{equation}
\ket{\uparrow}=
\begin{pmatrix}
1 \\ 0
\end{pmatrix},\qquad
\ket{\downarrow}=
\begin{pmatrix}
0 \\ 1
\end{pmatrix}
\end{equation}
as a basis in $\mathbb{C}^2$, we define $L$-operator
by assuming further that up and right arrows correspond
to spin up components while down and left arrows correspond
to spin down ones. For the $L$-operator we have then the expression
\begin{equation}\label{Lop}
L_{\alpha k}(\lambda_\alpha,\nu_k)=
\sin(\lambda_\alpha-\nu_k+\eta\,\sigma_\alpha^z \sigma_k^z )
+ \sin(2\eta)(\sigma_\alpha^{-}\sigma_k^{+}
+\sigma_\alpha^{+}\sigma_k^{-}),
\end{equation}
where $\sigma^z,\sigma^{\pm}=(1/2)(\sigma^x\pm\rmi\sigma^y)$
are Pauli matrices
and the subscripts in \eqref{Lop} indicate the space in which they act.

The monodromy matrix is an ordered product of $L$-operators.
We shall define it here as a product along a column, promoting
the corresponding vertical space $\mathcal{V}_\alpha$ to
be `auxiliary' space while the horizontal spaces
$\mathcal{H}_k$ will be treated as `quantum' spaces;
the space $\mathcal{H}=\ot_{k=1}^N\mathcal{H}_k$ is thus
the total quantum space. To define the monodromy matrix
it is more convenient to think of $L$-operator
as acting in $\mathcal{V}_\alpha\otimes\mathcal{H}$, moreover,
writing it as $2\times 2$ matrix in $\mathcal{V}_\alpha$
with quantum operator entries acting in $\mathcal{H}$,
\begin{equation}
L_{\alpha k}(\lambda_\alpha,\nu_k)=
\begin{pmatrix}
\sin(\lambda_\alpha-\nu_k+\eta\,\sigma_k^z) &  \sin(2\eta)\,\sigma_k^-\\
\sin(2\eta)\,\sigma_k^+ & \sin(\lambda_\alpha-\nu_k-\eta\,\sigma_k^z)
\end{pmatrix}_{[\alpha]}.
\end{equation}
Here $[\alpha]$ indicates that this is the matrix with respect to the
auxiliary space $\mathcal{V}_\alpha$ and
$\sigma_k^{\pm,z}$ stand for quantum operators in $\mathcal{H}$,
acting as Pauli matrices in $\mathcal{H}_k$ and identically elsewhere.
The monodromy matrix is defined by
\begin{equation}
T_\alpha(\lambda_\alpha)=
\prod_{k=1}^{\rtol{N}}
L_{\alpha k}(\lambda_\alpha,\nu_k)
=\begin{pmatrix}
A(\lambda_\alpha)& B(\lambda_\alpha) \\
C(\lambda_\alpha)& D(\lambda_\alpha)
\end{pmatrix}_{[\alpha]}.
\end{equation}
The operators $A(\lambda)=A(\lambda;\nu_1,\dots,\nu_N)$, etc,
act in $\mathcal{H}$ and play a fundamental role in QISM.

These operators, $A(\lambda)$, $B(\lambda)$, $C(\lambda)$, and
$D(\lambda)$ are subject to the Yang-Baxter algebra,
\begin{equation}\label{RTT}
R_{\alpha\alpha'}(\lambda,\lambda')
\big[T_\alpha(\lambda)\otimes T_{\alpha'}(\lambda')\big]=
\big[T_\alpha(\lambda')\otimes T_{\alpha'}(\lambda)\big]
R_{\alpha\alpha'}(\lambda,\lambda').
\end{equation}
generated by the six-vertex model $R$-matrix,
\begin{equation}
R_{\alpha\alpha'}(\lambda,\lambda')=
\begin{pmatrix}
f(\lambda',\lambda) & 0 &0 &0 \\
0 & g(\lambda',\lambda) &1 &0 \\
0 &1 & g(\lambda',\lambda) &0 \\
0 &0 &0 & f(\lambda',\lambda)
\end{pmatrix}_{[\alpha\alpha']}.
\end{equation}
where
\begin{equation}
f(\lambda',\lambda)=
\frac{\sin(\lambda-\lambda'+2\eta)}{\sin(\lambda-\lambda')},\qquad
g(\lambda',\lambda)=
\frac{\sin(2\eta)}{\sin(\lambda-\lambda')}.
\end{equation}
The relation \eqref{RTT} is also known as RTT relation and
it is a consequence of the similar RLL relation.
Among the sixteen relations contained in \eqref{RTT},
in particular, are the following
\begin{align} \label{BB}
B(\lambda)\, B(\lambda')&=
B(\lambda')\, B(\lambda),
\\ \label{AB}
A(\lambda)\, B(\lambda')&=
f(\lambda,\lambda')\, B(\lambda')\, A(\lambda)
+g(\lambda',\lambda)\, B(\lambda)\, A(\lambda'),
\\  \label{DB}
D(\lambda)\, B(\lambda')&=
f(\lambda',\lambda)\, B(\lambda')\, D(\lambda)
+g(\lambda,\lambda')\, B(\lambda)\, D(\lambda'),
\\  \label{CB}
C(\lambda)\, B(\lambda')&=
B(\lambda')\, C(\lambda) +
g(\lambda,\lambda')
\big[A(\lambda)\, D(\lambda')-A(\lambda')\,D(\lambda) \big],
\end{align}
which will be used below.

\begin{figure}[t]
\begin{center}
\psset{unit=.75cm,dotsep=.1,arrowscale=2}
\begin{pspicture}(0,-2)(2,7)
\psline(1,0)(1,2.2)
\psline(1,2.8)(1,6)
\psline[linestyle=dotted](1,2.3)(1,2.7)
\multips(0,1)(0,1){5}{\psline(.25,0)(1.75,0)}
\psline{->}(1,.6)(1,.7)\psline{->}(1,5.6)(1,5.7)
\rput(1,7){$\lambda$}
\rput(1,-1.5){$A(\lambda)$}
\end{pspicture}
\qquad
\begin{pspicture}(0,-2)(2,7)
\psline(1,0)(1,2.2)
\psline(1,2.8)(1,6)
\psline[linestyle=dotted](1,2.3)(1,2.7)
\multips(0,1)(0,1){5}{\psline(.25,0)(1.75,0)}
\psline{->}(1,.6)(1,.7)\psline{<-}(1,5.3)(1,5.4)
\rput(1,7){$\lambda$}
\rput(1,-1.5){$B(\lambda)$}
\end{pspicture}
\qquad
\begin{pspicture}(0,-2)(2,7)
\psline(1,0)(1,2.2)
\psline(1,2.8)(1,6)
\psline[linestyle=dotted](1,2.3)(1,2.7)
\multips(0,1)(0,1){5}{\psline(.25,0)(1.75,0)}
\psline{<-}(1,.3)(1,.4)\psline{->}(1,5.6)(1,5.7)
\rput(1,7){$\lambda$}
\rput(1,-1.5){$C(\lambda)$}
\end{pspicture}
\qquad
\begin{pspicture}(0,-2)(2,7)
\psline(1,0)(1,2.2)
\psline(1,2.8)(1,6)
\psline[linestyle=dotted](1,2.3)(1,2.7)
\multips(0,1)(0,1){5}{\psline(.25,0)(1.75,0)}
\psline{<-}(1,.3)(1,.4)\psline{<-}(1,5.3)(1,5.4)
\rput(1,7){$\lambda$}
\rput(1,-1.5){$D(\lambda)$}
\end{pspicture}
\end{center}
\caption{Graphical interpretation of the
operators $A(\lambda)$, $B(\lambda)$, $C(\lambda)$, and $D(\lambda)$.}
\label{fig-abcd}
\end{figure}
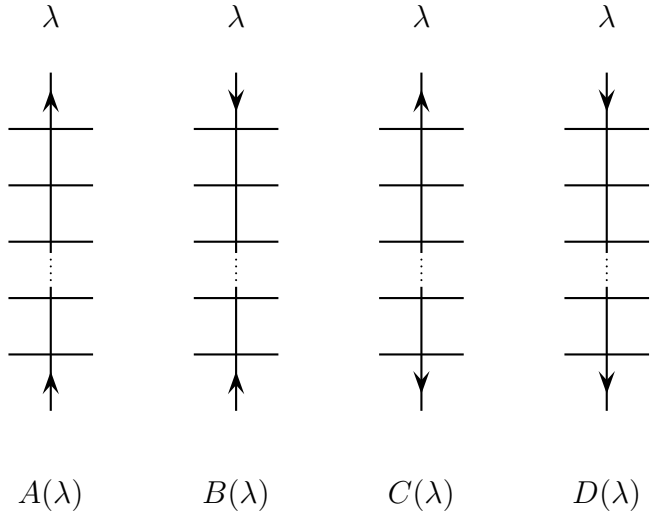

The operators $A(\lambda)$, $B(\lambda)$, $C(\lambda)$, and $D(\lambda)$
admit simple graphical
interpretation as columns of the lattice, with top and bottom
arrows fixed, see figure \ref{fig-abcd}.
Taking into account this nice graphical interpretation
and exploiting further the correspondence between arrows and
spin states to express
arrows on the left and right boundaries of the lattice
in terms of `all spins down' and `all spins up' states,
\begin{equation}
\bra{\Downarrow}=\ot_{k=1}^N\ {}_k\bra{\downarrow}\,,\qquad
\ket{\Uparrow}=\ot_{k=1}^{N}\ \ket{\uparrow}_k,
\end{equation}
where $\ket{\uparrow}_k$ and $\ket{\downarrow}_k$
are the spin up and spin down vectors of the space $\mathcal{H}_k$,
it can be easily seen that the
partition function, $Z_N=Z_N(\{\lambda_\alpha\};\{\nu_k\})$,
of the six-vertex model with DWBC is given by
\begin{equation}\label{ZBBB}
Z_N= \bra{\Downarrow} B(\lambda_N) \cdots B(\lambda_1)\ket{\Uparrow}.
\end{equation}
Note, that because of  relation \eqref{BB}
the actual order of operators
$B(\lambda)$ is not important in this formula.

The explicit expression for the partition function,
found and proven in \cite{I-87}, see also \cite{ICK-92}, reads
\begin{equation}\label{ZN}
Z_N=
\frac{\prod_{\alpha=1}^N \prod_{k=1}^N
a(\lambda_\alpha,\nu_k)b(\lambda_\alpha,\nu_k)}{
\prod_{1\leq\alpha<\beta\leq N}d(\lambda_\beta,\lambda_\alpha)
\prod_{1\leq j<k\leq N}d(\nu_j,\nu_k)}\;
{\det}_N^{} T
\end{equation}
where
\begin{equation}\label{d}
d(\lambda,\lambda'):=\sin(\lambda-\lambda')
\end{equation}
and the functions $a(\lambda,\nu)$ and $b(\lambda,\nu)$ are defined
in \eqref{abc}. The $N\times N$ matrix $T$ is given by
\begin{equation}\label{matT}
T_{\alpha k}=t(\lambda_\alpha,\nu_k),\qquad
t(\lambda,\nu)=\frac{\sin(2\eta)}{\sin(\lambda-\nu+\eta)
\sin(\lambda-\nu-\eta)}.
\end{equation}
In \cite{I-87,ICK-92} the determinant formula \eqref{ZN}
was proven on the basis
of certain recursion formulae, established in \cite{K-82}.
Another proof, solely on the basis of Yang-Baxter algebra,
was given in \cite{BPZ-02};
in the next section we sketch the derivation.

In the homogenous limit, when $\lambda_\alpha=\lambda$ and
$\nu_k=0$, expression \eqref{ZN}
turns into the following one \cite{ICK-92}
\begin{equation}\label{ZNhom}
Z_N=\frac{[\sin(\lambda-\eta)\sin(\lambda+\eta)]^{N^2}}
{\prod_{n=1}^{N-1}(n!)^2}\;
{\det}_N^{} \varPhi
\end{equation}
where
\begin{equation}\label{varphi}
\varPhi_{\alpha k}=\partial_{\lambda}^{\alpha+k-2}
\varphi(\lambda,\eta),\qquad
\varphi(\lambda,\eta)=
\frac{\sin(2\eta)}{\sin(\lambda-\eta)\sin(\lambda+\eta)}.
\end{equation}
The procedure of obtaining \eqref{ZNhom} from \eqref{ZN}
was explained in detail in \cite{ICK-92}.
In our treatment of the correlation functions we shall proceed
in the same way, first obtaining an expression for the
inhomogeneous model, and next taking the homogenous limit.

\section{One-point boundary correlation functions}

Here we recall the main results of paper \cite{BPZ-02}
on one-point boundary correlation functions. In the next section
we shall explain how this approach can be used to
compute two-point boundary correlation functions.

\begin{figure}[t]
\begin{center}
\psset{unit=.75cm,dotsep=.1,arrowscale=2,dotscale=1.5}
\begin{pspicture}(0,-1)(9,10)
\multips(1,0)(1,0){8}{\psline(0,0)(0,3.2)}
\multips(0,1)(0,1){8}{\psline(0,0)(2.2,0)}
\multips(3,1)(0,1){8}{\psline(-.2,0)(3.2,0)}
\multips(1,9)(1,0){8}{\psline(0,0)(0,-5.2)}
\multips(9,1)(0,1){8}{\psline(0,0)(-2.2,0)}
\multips(1,3.3)(1,0){8}{\psline[linestyle=dotted](0,0)(0,.4)}
\multips(2.3,1)(0,1){8}{\psline[linestyle=dotted](0,0)(.4,0)}
\multips(6.3,1)(0,1){8}{\psline[linestyle=dotted](0,0)(.4,0)}
\multirput(1,.25)(1,0){8}{\psline{->}(0,0)(0,.5)}
\multirput(.25,1)(0,1){8}{\psline{<-}(0,0)(.5,0)}
\multirput(1,8.75)(1,0){8}{\psline{->}(0,0)(0,-.5)}
\multirput(8.75,1)(0,1){8}{\psline{<-}(0,0)(-.5,0)}
\rput(5,7.25){\psline{->}(0,0)(0,.5)}
\rput(3.25,8){\psline{<-}(0,0)(.5,0)}
\rput(1.25,8){\psline{<-}(0,0)(.5,0)}
\rput(4.25,8){\psline{<-}(0,0)(.5,0)}
\rput(5.25,8){\psline{->}(0,0)(.5,0)}
\rput(7.25,8){\psline{->}(0,0)(.5,0)}
\multirput(1,7.75)(1,0){4}{\psline{->}(0,0)(0,-.5)}
\multirput(6,7.75)(1,0){3}{\psline{->}(0,0)(0,-.5)}
\pscircle(5,8){.3}
\rput(5.1,10){$r$}
\rput(8,10){$1$}
\psline{->}(7.5,10)(5.5,10)
\rput(4.5,-1){(a)}
\end{pspicture}
\qquad
\begin{pspicture}(0,-1)(9,10)
\multips(1,0)(1,0){8}{\psline(0,0)(0,3.2)}
\multips(0,1)(0,1){8}{\psline(0,0)(2.2,0)}
\multips(3,1)(0,1){8}{\psline(-.2,0)(3.2,0)}
\multips(1,9)(1,0){8}{\psline(0,0)(0,-5.2)}
\multips(9,1)(0,1){8}{\psline(0,0)(-2.2,0)}
\multips(1,3.3)(1,0){8}{\psline[linestyle=dotted](0,0)(0,.4)}
\multips(2.3,1)(0,1){8}{\psline[linestyle=dotted](0,0)(.4,0)}
\multips(6.3,1)(0,1){8}{\psline[linestyle=dotted](0,0)(.4,0)}
\multirput(1,.25)(1,0){8}{\psline{->}(0,0)(0,.5)}
\multirput(.25,1)(0,1){8}{\psline{<-}(0,0)(.5,0)}
\multirput(1,8.75)(1,0){8}{\psline{->}(0,0)(0,-.5)}
\multirput(8.75,1)(0,1){8}{\psline{<-}(0,0)(-.5,0)}
\rput(3.25,8){\psline{<-}(0,0)(.5,0)}
\rput(1.25,8){\psline{<-}(0,0)(.5,0)}
\rput(4.25,8){\psline{<-}(0,0)(.5,0)}
\multirput(1,7.75)(1,0){4}{\psline{->}(0,0)(0,-.5)}
\pscircle(4.5,8){.3}
\rput(5.1,10){$r$}
\rput(8,10){$1$}
\psline{->}(7.5,10)(5.5,10)
\rput(4.5,-1){(b)}
\end{pspicture}
\end{center}
\caption{Boundary
one-point correlation functions: (a) function $H_N^{(r)}$;
(b) function $G_N^{(r)}$.}
\label{fig-1cor}
\end{figure}
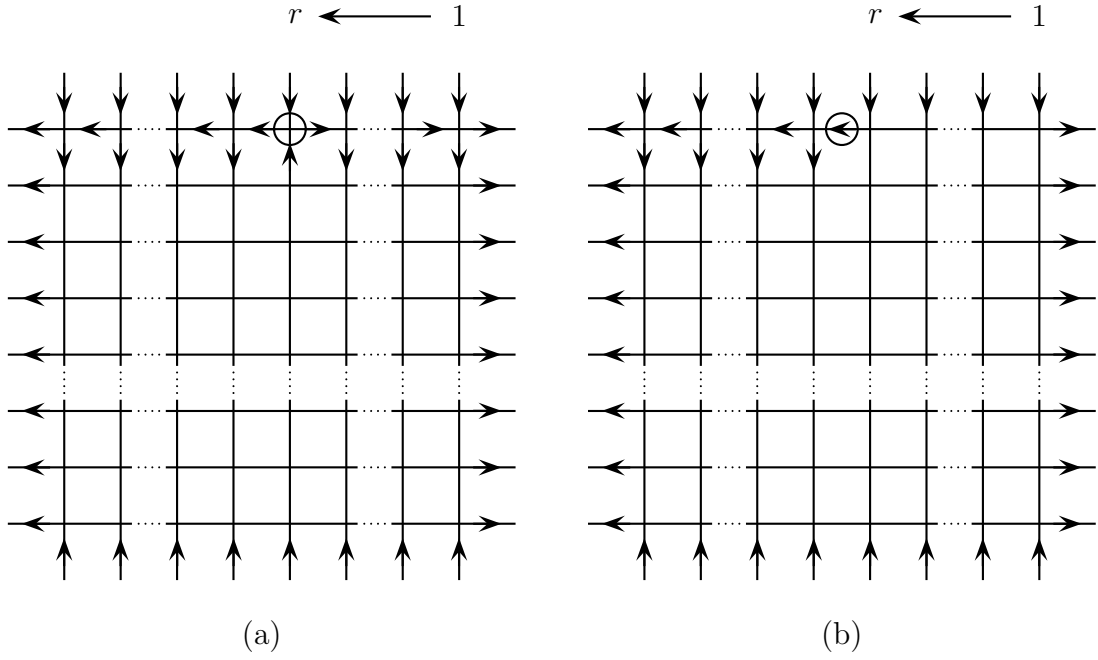

In paper \cite{BPZ-02} two closely related kinds of one-point boundary
correlation functions were considered.
The first correlation function,
denoted as $H_N^{(r)}$, is the
probability of finding the $r$-th vertex (counted from the right)
on the first row in the state $i=5$.
Formally, this correlation function can be defined by
\begin{equation}\label{H_Nr}
H_N^{(r)}=Z_N^{-1}
\bra{\Downarrow} B(\lambda_N)\dots B(\lambda_{r+1})\, q_1\,
B(\lambda_r)\, p_1\, B(\lambda_{r-1})
\cdots B(\lambda_1)\ket{\Uparrow}
\end{equation}
where  $q_k$ and $p_k$ denote
projection operators on the spin up and spin down states, at $k$-th
`site', respectively,
\begin{equation}
q_k^{}=\frac{1}{2}(1-\sigma_k^z),\qquad
p_k^{}=\frac{1}{2}(1+\sigma_k^z).
\end{equation}
The second correlation, denoted as $G_N^{(r)}$,
is the boundary `polarization', i.e. the probability
of finding an arrow pointing  left on the horizontal
edge of the first row  between $r$-th and $(r+1)$-th columns.
One can define this correlation function by
\begin{equation}\label{G_Nr}
G_N^{(r)}=Z_N^{-1}
\bra{\Downarrow} B(\lambda_N)\dots B(\lambda_{r+1})\, q_1\,
B(\lambda_r) \cdots B(\lambda_1)\ket{\Uparrow}.
\end{equation}
Due to DWBC, the two correlation function $H_N^{(r)}$ and $G_N^{(r)}$
are related by
\begin{equation}\label{GHHG}
G_N^{(r)}=\sum_{\alpha=1}^{r}H_N^{(\alpha)};\qquad
H_N^{(r)}=G_N^{(r)}-G_N^{(r-1)}.
\end{equation}
These relations can be easily found by exploring the graphical
interpretation of these functions, see figure \ref{fig-1cor}.

To compute the one-point boundary correlation functions
one can introduce
the operators $\wt A(\lambda)$, $\wt B(\lambda)$, $\wt C(\lambda)$ and
$\wt D(\lambda)$ as entries
of the monodromy matrix on $N-1$ sites,
\begin{equation}
\wt T_\alpha(\lambda_\alpha):=
\prod_{k=2}^{\rtol{N}}
L_{\alpha,k}(\lambda_\alpha,\nu_k)
=
\begin{pmatrix}
\wt A(\lambda_\alpha) & \wt B(\lambda_\alpha)\\
\wt C(\lambda_\alpha)& \wt D(\lambda_\alpha)
\end{pmatrix}.
\end{equation}
Correspondingly, one can define
\begin{equation}
\bra{\wt\Downarrow}=\ot_{k=2}^{N}\ {}_k\bra{\uparrow}\,,\qquad
\ket{\wt\Uparrow}=\ot_{k=2}^{N}\ \ket{\uparrow}_k^{}\;.
\end{equation}
One has
\begin{equation}
\wt A(\lambda)\ket{\wt\Uparrow}
=\prod_{k=2}^{N} a(\lambda,\nu_k)\ket{\wt\Uparrow},\qquad
\wt D(\lambda)\ket{\wt\Uparrow}
=\prod_{k=2}^{N} b(\lambda,\nu_k)\ket{\wt\Uparrow}.
\end{equation}
In what follows, to simplify writing,
we shall omit tildes over operators and vectors.

The role of these new operators in application to the
one-point boundary correlators is quite clear from
their graphical interpretation shown on figure \ref{fig-abcd}
and the analogous graphical representation for
the correlation functions shown on figure
\ref{fig-1cor}. Specifically, in the case of $H_N^{(r)}$ all
vertices of the first row are fixed, thus leading to the following
formula in terms of the operators on $N-1$ sites
\begin{multline}\label{HNrBAB}
H_N^{(r)}= Z_N^{-1}\, c
\prod_{\alpha=r+1}^{N} a(\lambda_\alpha,\nu_1)\,
\prod_{\alpha=1}^{r-1} b(\lambda_\alpha,\nu_1)
\\ \times
\bra{\Downarrow}
B(\lambda_N)\cdots
B(\lambda_{r+1})\, A(\lambda_{r})\,
B(\lambda_{r-1})\cdots B(\lambda_{1})
\ket{\Uparrow}.
\end{multline}
Similarly, the correlation function
$G_N^{(r)}$ can be written as a sum of $r$ such terms
\begin{multline}\label{GNrBAB}
G_N^{(r)}= Z_N^{-1}\, c
\sum_{\alpha=1}^{r}
\prod_{\beta=\alpha+1}^{N} a(\lambda_\beta,\nu_1)\,
\prod_{\beta=1}^{\alpha-1} b(\lambda_\beta,\nu_1)
\\ \times
\bra{\Downarrow}
B(\lambda_N)\cdots
B(\lambda_{\alpha+1})\, A(\lambda_{\alpha})\,
B(\lambda_{\alpha-1})\cdots B(\lambda_{1})
\ket{\Uparrow}.
\end{multline}

On the first stage of computation the correlation function are
expressed in terms of the partition functions on  $(N-1)\times (N-1)$
sublattices. Using the commutation relation \eqref{AB} we have
\begin{equation}
A(\lambda_r) \prod_{\alpha=1}^{r-1} B(\lambda_\alpha)\ket{\Uparrow}
=\sum_{\alpha=1}^{r}
\prod_{k=2}^{N}a(\lambda_\alpha,\nu_k)
\frac{g(\lambda_\alpha,\lambda_r)}{f(\lambda_\alpha,\lambda_r)}
\prod_{\substack{\beta=1\\ \beta\ne\alpha}}^{r}
f(\lambda_\alpha,\lambda_\beta)
\prod_{\substack{\beta=1\\ \beta\ne\alpha}}^{r}
B(\lambda_\beta)\ket{\Uparrow}.
\end{equation}
The last relation allows one to obtain, in the case of
$H_N^{(r)}$, the expression
\begin{multline}\label{HNrZ}
H_N^{(r)}= Z_N^{-1}\, c
\prod_{\alpha=r+1}^{N} a(\lambda_\alpha,\nu_1)\,
\prod_{\alpha=1}^{r-1} b(\lambda_\alpha,\nu_1)
\\ \times
\sum_{\beta=1}^{r} \prod_{k=2}^{N} a(\lambda_\beta,\nu_k)
\frac{g(\lambda_\beta,\lambda_r)}{f(\lambda_\beta,\lambda_r)}
\prod_{\substack{\gamma=1\\ \gamma\ne\beta}}^{r}
f(\lambda_\beta,\lambda_\gamma)
Z_{N-1}\big(\{\lambda_\delta\}_{\delta=1,\delta\ne\alpha}^N;
\{\nu_k\}_{k=2}^N\big)
\end{multline}
The corresponding expression for $G_N^{(r)}$ can be immediately
obtained using the relationship with the function
$H_N^{(r)}$, see \eqref{GHHG}. However, as explained in \cite{BPZ-02},
in this case one gets the result in terms of some double sum
which is not actually an analogue of \eqref{HNrZ}. It was
pointed out that the analogue of \eqref{HNrZ} can be found
if one uses instead the relation
\begin{multline}
\sum_{\alpha=1}^{r}
\prod_{\beta=\alpha+1}^{r}
a(\lambda_\beta,\nu_1)
\prod_{\beta=1}^{\alpha-1}
b(\lambda_\beta,\nu_1)\;
B(\lambda_r)\cdots B(\lambda_{\alpha+1})
A(\lambda_\alpha)
B(\lambda_{\alpha-1})
\cdots
B(\lambda_1)
\ket{\Uparrow}
\\
=
\sum_{\alpha=1}^{r}
\prod_{k=2}^{N}a(\lambda_\alpha,\nu_k)
\prod_{\substack{\beta=1 \\ \beta\ne\alpha}}^{r} b(\lambda_\beta,\nu_1)
\prod_{\substack{\beta=1\\ \beta\ne\alpha}}^{r}
f (\lambda_\alpha,\lambda_\beta)
\prod_{\substack{\beta=1\\ \beta\ne\alpha}}^{r} B(\lambda_\beta)
\ket{\Uparrow}
\end{multline}
which can be obtained by means of \eqref{AB} by taking into
account that LHS here is symmetric with respect to permutations of
$\lambda_1,\dots,\lambda_r$ by
construction (for more details on derivation of such relations see,
e.g., \cite{KBI-93}, sections VII.2 and XXII.2).
In this way one obtains
\begin{multline}\label{GNrZ}
G_N^{(r)}
=Z_N^{-1} \prod_{\alpha=r+1}^{N} a(\lambda_\alpha,\nu_1)\,
 \prod_{\alpha=1}^{r} b(\lambda_\alpha,\nu_1)
\\ \times
\sum_{\beta=1}^{r}
\frac{c}{b(\lambda_\beta,\nu_1)}
\prod_{k=2}^{N}a(\lambda_\beta,\nu_k)
\prod_{\substack{\gamma=1\\ \gamma\ne\beta}}^{r}
f (\lambda_\beta,\lambda_\gamma)
Z_{N-1}\big(\{\lambda_\delta\}_{\delta=1,\delta\ne\beta}^N;
\{\nu_k\}_{k=2}^N\big).
\end{multline}
As explained in paper \cite{BPZ-02} this representation
is important since it allows one to prove the determinant formula
\eqref{ZN}. Indeed, since by definition $G_N^{(N)}=1$,  expression
\eqref{GNrZ} turns into some recurrence relation connecting
the partition functions $Z_{N}$ and $Z_{N-1}$;
note, that here all the parameters
in the sets $\{\lambda_\alpha\}_{\alpha=1}^N$ and $\{\nu_k\}_{k=1}^N$
are assumed to be completely arbitrary
(cf. \cite{I-87,ICK-92}).
As explained in \cite{BPZ-02}, it can
be readily shown, in virtue of the Kramer rule
and of some particular identity, that the solution of this recurrence
relation, with the initial condition
$Z_1=c$, is given by determinant formula \eqref{ZN}.

Thus, the expression for the partition function being proved
within the considered framework, it can be used to obtain
similar representations for one-point functions.
Substituting the expression
for $Z_{N-1}$ in \eqref{HNrZ} and \eqref{GNrZ} gives rise to the
determinant formulae for one-point functions, see
again \cite{BPZ-02} for details. We end up this section
by quoting the results.

The function $H_N^{(r)}$ is given by
\begin{equation}
H_N^{(r)} = \frac{c\prod_{k=2}^{N}d(\nu_1,\nu_k)}{
\prod_{\alpha=1}^{r}a(\lambda_\alpha,\nu_1)
\prod_{\alpha=r}^{N}b(\lambda_\alpha,\nu_1)}\;
\frac{{\det}_N^{} V}{{\det}_N^{} T}
\end{equation}
where the matrix $V$ differs from the matrix $T$, equation \eqref{matT},
just by the elements of the
first column,
\begin{equation}
V_{\alpha,1}=v_r(\lambda_\alpha);\qquad
V_{\alpha,k}=T_{\alpha,k},\quad k=2,\dots,N.
\end{equation}
Here the function $v_r(\lambda)$ is given by
\begin{equation}\label{vr}
v_r(\lambda)= \frac{\prod_{\alpha=r+1}^{N} d(\lambda_\alpha,\lambda)
\prod_{\alpha=1}^{r-1} e(\lambda_\alpha,\lambda)
}{\prod_{k=2}^{N}b(\lambda,\nu_k)}
\end{equation}
where
\begin{equation}
e(\lambda,\lambda')=\sin(\lambda-\lambda'+2\eta)
\end{equation}
and the function $d(\lambda,\lambda')$ is defined in \eqref{d}.

The function $G_N^{(r)}$ is given by
\begin{equation}
G_N^{(r)} = \frac{\prod_{k=2}^{N}d(\nu_1,\nu_k)}{
\prod_{\alpha=1}^{r}a(\lambda_\alpha,\nu_1)
\prod_{\alpha=r+1}^{N}b(\lambda_\alpha,\nu_1)}\;
\frac{{\det}_N^{} S}{{\det}_N^{} T}
\end{equation}
where the matrix $S$ also differs from $T$ just by the elements of the
first column,
\begin{equation}
S_{\alpha,1}=s_r(\lambda_\alpha);\qquad
S_{\alpha,k}=T_{\alpha,k},\quad k=2,\dots,N.
\end{equation}
Here the function $s_r(\lambda)$ is given by
\begin{equation}
s_r(\lambda)= \frac{\prod_{\alpha=r+1}^{N} d(\lambda_\alpha,\lambda)
\prod_{\alpha=1}^{r} e(\lambda_\alpha,\lambda)
}{\prod_{k=1}^{N}b(\lambda,\nu_k)}.
\end{equation}

In the homogeneous limit the following formulae are valid.
The function $H_N^{(r)}$ is given by
\begin{equation}\label{Hhom}
H_N^{(r)}=\frac{(N-1)!\,\sin(2\eta)}
{\big[\sin(\lambda+\eta)\big]^r\big[\sin(\lambda-\eta)\big]^{N-r+1}}\;
\frac{{\det}_N \varPsi}{{\det}_N \varPhi}
\end{equation}
where the matrix $\varPsi$ differs from the matrix $\varPhi$,
equation \eqref{varphi}, just by the elements of the last column
\begin{equation}
\varPsi_{\alpha,k}=\varPhi_{\alpha,k},\quad k=1,\dots,N-1;\qquad
\varPsi_{\alpha,N}=\partial_\eps^{\alpha-1}
\frac{(\sin\eps)^{N-r}[\sin(\eps-2\eta)]^{r-1}}
{[\sin(\eps+\lambda-\eta)]^{N-1}}\bigg|_{\eps=0}.
\end{equation}
Similarly, the function $G_N^{(r)}$ is given by
\begin{equation}\label{Ghom}
G_N^{(r)}=\frac{(N-1)!}
{\big[\sin(\lambda+\eta)\big]^r\big[\sin(\lambda-\eta)\big]^{N-r}}\;
\frac{{\det}_N \varTheta}{{\det}_N \varPhi}
\end{equation}
where
\begin{equation}
\varTheta_{\alpha,k}=\varPhi_{\alpha,k},\quad k=1,\dots,N-1;\qquad
\varTheta_{\alpha,N}=-\partial_\eps^{\alpha-1}
\frac{(\sin\eps)^{N-r}[\sin(\eps-2\eta)]^{r}}
{[\sin(\eps+\lambda-\eta)]^{N}}\bigg|_{\eps=0}.
\end{equation}

\section{Two-point boundary correlations}\label{sec-twop}

\begin{figure}[t]
\begin{center}
\psset{unit=.75cm,dotsep=.1,arrowscale=2,dotscale=1.5}
\begin{pspicture}(0,-1)(9,10)
\multips(1,0)(1,0){8}{\psline(0,0)(0,3.2)}
\multips(0,1)(0,1){8}{\psline(0,0)(1.2,0)}
\multips(2,1)(0,1){8}{\psline(-.2,0)(2.2,0)}
\multips(5,1)(0,1){8}{\psline(-.2,0)(2.2,0)}
\multips(1,9)(1,0){8}{\psline(0,0)(0,-5.2)}
\multips(9,1)(0,1){8}{\psline(0,0)(-1.2,0)}
\multips(1,3.3)(1,0){8}{\psline[linestyle=dotted](0,0)(0,.4)}
\multips(1.3,1)(0,1){8}{\psline[linestyle=dotted](0,0)(.4,0)}
\multips(4.3,1)(0,1){8}{\psline[linestyle=dotted](0,0)(.4,0)}
\multips(7.3,1)(0,1){8}{\psline[linestyle=dotted](0,0)(.4,0)}
\multirput(1,.25)(1,0){8}{\psline{->}(0,0)(0,.5)}
\multirput(.25,1)(0,1){8}{\psline{<-}(0,0)(.5,0)}
\multirput(1,8.75)(1,0){8}{\psline{->}(0,0)(0,-.5)}
\multirput(8.75,1)(0,1){8}{\psline{<-}(0,0)(-.5,0)}
\rput(6,7.25){\psline{->}(0,0)(0,.5)}
\rput(3.25,8){\psline{<-}(0,0)(.5,0)}
\rput(2.25,8){\psline{<-}(0,0)(.5,0)}
\rput(5.25,8){\psline{<-}(0,0)(.5,0)}
\rput(6.25,8){\psline{->}(0,0)(.5,0)}
\multirput(1,7.75)(1,0){5}{\psline{->}(0,0)(0,-.5)}
\multirput(7,7.75)(1,0){2}{\psline{->}(0,0)(0,-.5)}
\rput(3,1.25){\psline{<-}(0,0)(0,.5)}
\rput(2.25,1){\psline{<-}(0,0)(.5,0)}
\rput(3.25,1){\psline{->}(0,0)(.5,0)}
\rput(5.25,1){\psline{->}(0,0)(.5,0)}
\rput(6.25,1){\psline{->}(0,0)(.5,0)}
\multirput(1,1.75)(1,0){2}{\psline{<-}(0,0)(0,-.5)}
\multirput(4,1.75)(1,0){5}{\psline{<-}(0,0)(0,-.5)}
\pscircle(6,8){.3}
\pscircle(3,1){.3}
\rput(6.1,10){$r_1$}
\rput(8,10){$1$}
\psline{->}(7.5,10)(6.5,10)
\rput(3.1,-1){$r_2$}
\rput(8,-1){$1$}
\psline{->}(7.5,-1)(3.5,-1)
\end{pspicture}
\end{center}
\caption{The two-point correlation function $H_N^{(r_1,r_2)}$.}
\label{fig-2cor}
\end{figure}
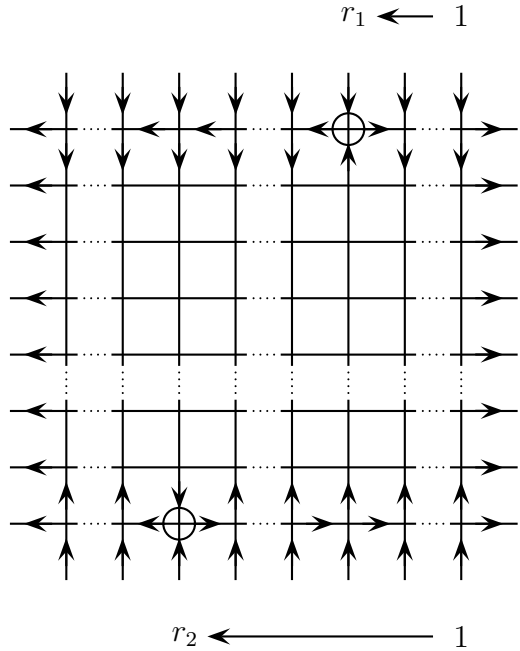

The one-point boundary correlation functions just considered
can be directly generalized to the case of two-point ones
\cite{FP-04}.
Here we shall consider in detail derivation of the function
$H_N^{(r_1,r_2)}$, which gives the probability of finding vertices
of type $i=5$ on the opposite, top and bottom, boundaries.
More precisely, we define $H_N^{(r_1,r_2)}$
as the  probability to find vertices of type $i=5$
both at $r_1$-th position (counted from the right) of the first row
and at $r_2$-th position of the last row, see figure \ref{fig-2cor}.
Formally, if $r_1<r_2$ then one can define this correlation
function by
\begin{multline}
H_N^{(r_1,r_2)}= Z_N^{-1}
\bra{\Downarrow}
B(\lambda_N)\cdots
B(\lambda_{r_2+1})\, q_N^{}\, B(\lambda_{r_2})\, p_N^{}\,
 B(\lambda_{r_2-1})\cdots
\\
\times \cdots
B(\lambda_{r_1+1})\, q_1^{}\, B(\lambda_{r_1})\, p_1^{}\,
B(\lambda_{r_1-1})\cdots B(\lambda_{1})
\ket{\Uparrow}.
\end{multline}
For $r_1=r_2$ or $r_1>r_2$ one has  analogous expressions.

Quite similarly, one can also consider the function $G_N^{(r_1,r_2)}$,
giving the probability of finding arrows pointing  left
on the first and the last rows.
The functions $H_N^{(r_1,r_2)}$
and $G_N^{(r_1,r_2)}$ are related to each other
just as in the one-point case;  here we focus on
the function $H_N^{(r_1,r_2)}$ since it is the most interesting
for combinatorial applications. The result for
$G_N^{(r_1,r_2)}$ can be found using its obvious relation with
$H_N^{(r_1,r_2)}$.

To compute the function $H_N^{(r_1,r_2)}$ we introduce
the operators $\wt A(\lambda)$, $\wt B(\lambda)$, $\wt C(\lambda)$ and
$\wt D(\lambda)$ as entries
of the monodromy matrix on $N-2$ sites,
\begin{equation}
\wt T_\alpha(\lambda_\alpha):=
\prod_{k=2}^{\rtol{N-1}}
L_{\alpha,k}(\lambda_\alpha,\nu_k)
=
\begin{pmatrix}
\wt A(\lambda_\alpha) & \wt B(\lambda_\alpha)\\
\wt C(\lambda_\alpha)& \wt D(\lambda_\alpha)
\end{pmatrix}.
\end{equation}
Correspondingly, we define
\begin{equation}
\bra{\wt\Downarrow}=\ot_{k=2}^{N-1}\ {}_k\bra{\uparrow}\,,\qquad
\ket{\wt\Uparrow}=\ot_{k=2}^{N-1}\ \ket{\uparrow}_k^{}\;.
\end{equation}
We have
\begin{equation}
\wt A(\lambda)\ket{\wt\Uparrow}
=\prod_{k=2}^{N-1} a(\lambda,\nu_k)\ket{\wt\Uparrow},\qquad
\wt D(\lambda)\ket{\wt\Uparrow}
=\prod_{k=2}^{N-1} b(\lambda,\nu_k)\ket{\wt\Uparrow}.
\end{equation}
In what follows, to simplify writing,
we shall again omit tildes over operators and vectors.

On the first stage of computation we express the correlation function
$H_N^{(r_1,r_2)}$ in terms of partition functions on
$(N-2)\times(N-2)$ lattices. This can be done
by directly following the procedure
of the previous section. For an alternative
way of performing
this preliminary step, see also \cite{FP-04}.

Let us assume that $r_1\ne r_2$. If $r_1< r_2$ then,
using the same arguments as for one-point correlations
(e.g., by invoking the graphical interpretation of the operators
shown on figure \ref{fig-abcd}), for the two-point correlator
$H_N^{(r_1,r_2)}$ we write
\begin{multline}\label{HNrrBB}
H_N^{(r_1,r_2)}= Z_N^{-1}\, c^2
\prod_{\alpha=r_1+1}^{N} a(\lambda_\alpha,\nu_1)\,
\prod_{\alpha=1}^{r_1-1} b(\lambda_\alpha,\nu_1)
\prod_{\alpha=r_2+1}^{N} b(\lambda_\alpha,\nu_N)\,
\prod_{\alpha=1}^{r_2-1} a(\lambda_\alpha,\nu_N)
\\ \times
\bra{\Downarrow}
B(\lambda_N)\cdots
B(\lambda_{r_2+1})\, D(\lambda_{r_2})\,
B(\lambda_{r_2-1})\cdots
B(\lambda_{r_1+1})\, A(\lambda_{r_1})\,
B(\lambda_{r_1-1})\cdots B(\lambda_{1})
\ket{\Uparrow}.
\end{multline}
Acting first with the operator $A(\lambda_{r_1})$ on the right, using
\begin{equation}
A(\lambda_r) \prod_{\alpha=1}^{r-1} B(\lambda_\alpha)\ket{\Uparrow}
=\sum_{\alpha=1}^{r}
\prod_{k=2}^{N-1}a(\lambda_\alpha,\nu_k)
\frac{g(\lambda_\alpha,\lambda_r)}{f(\lambda_\alpha,\lambda_r)}
\prod_{\substack{\beta=1\\ \beta\ne\alpha}}^{r}
f(\lambda_\alpha,\lambda_\beta)
\prod_{\substack{\beta=1\\ \beta\ne\alpha}}^{r}
B(\lambda_\beta)\ket{\Uparrow},
\end{equation}
and next with the operator $D(\lambda_{r_2})$, using
\begin{equation}
D(\lambda_r) \prod_{\alpha=1}^{r-1} B(\lambda_\alpha)\ket{\Uparrow}
=\sum_{\alpha=1}^{r}
\prod_{k=2}^{N-1}b(\lambda_\alpha,\nu_k)
\frac{g(\lambda_r,\lambda_\alpha)}{f(\lambda_r,\lambda_\alpha)}
\prod_{\substack{\beta=1\\ \beta\ne\alpha}}^{r}
f(\lambda_\beta,\lambda_\alpha)
\prod_{\substack{\beta=1\\ \beta\ne\alpha}}^{r}
B(\lambda_\beta)\ket{\Uparrow},
\end{equation}
we straightforwardly obtain
\begin{multline}\label{HNrrZ}
H_N^{(r_1,r_2)}= Z_N^{-1}\; c^2
\prod_{\alpha=r_1+1}^{N} a(\lambda_\alpha,\nu_1)\,
\prod_{\alpha=1}^{r_1-1} b(\lambda_\alpha,\nu_1)
\prod_{\alpha=r_2+1}^{N} b(\lambda_\alpha,\nu_N)\,
\prod_{\alpha=1}^{r_2-1} a(\lambda_\alpha,\nu_N)
\\ \times
\sum_{\alpha=1}^{r_1}
\sum_{\substack{\beta=1\\ \beta\ne\alpha}}^{r_2}
\prod_{k=2}^{N-1} a(\lambda_\alpha,\nu_k)
\frac{g(\lambda_\alpha,\lambda_{r_1})}{f(\lambda_\alpha,\lambda_{r_1})}
\prod_{\substack{\gamma=1\\\gamma\ne\alpha}}^{r_1}
f(\lambda_\alpha,\lambda_\gamma)
\prod_{k=2}^{N-1} b(\lambda_\beta,\nu_k)
\frac{g(\lambda_{r_2},\lambda_\beta)}{f(\lambda_{r_2},\lambda_\beta)}
\prod_{\substack{\gamma=1\\ \gamma\ne\alpha,\beta}}^{r_2}
f(\lambda_\gamma,\lambda_\beta)
\\ \times
Z_{N-2}\Big(\{\lambda_\delta\}_{\delta=1,\delta\ne\alpha,\beta}^N;
\{\nu_k\}_{k=2}^{N-1}\Big).
\end{multline}
If $r_1>r_2$ then one should first
act with the operator $D(\lambda_{r_2})$,
and next with the operator $A(\lambda_{r_1})$.
It can be easily verified that one obtains exactly the
same expression, i.e., formula \eqref{HNrrZ} is valid for
$r_1\ne r_2$.

Consider now the case $r_1=r_2=r$.
In this case instead of \eqref{HNrrBB} we have
\begin{multline}\label{HNrrC}
H_N^{(r,r)}= Z_N^{-1}\, c^2
\prod_{\alpha=r+1}^{N} a(\lambda_\alpha,\nu_1)\, b(\lambda_\alpha,\nu_N)
\prod_{\alpha=1}^{r-1} b(\lambda_\alpha,\nu_1)\, a(\lambda_\alpha,\nu_N)
\\ \times
\bra{\Downarrow}
B(\lambda_N)\cdots
B(\lambda_{r+1})\, C(\lambda_r)\,
B(\lambda_{r-1})\cdots
B(\lambda_{1})
\ket{\Uparrow}.
\end{multline}
Using commutation relation \eqref{CB} the following formula
can be derived (see \cite{KBI-93}, section VII.2.2)
\begin{multline}
C(\lambda_r) \prod_{\alpha=1}^{r-1} B(\lambda_\alpha)\ket{\Uparrow}
=\sum_{\substack{\alpha,\beta=1\\ \alpha\ne\beta}}^{r}
\prod_{k=2}^{N-1}a(\lambda_\alpha,\nu_k)
\prod_{k=2}^{N-1}b(\lambda_\beta,\nu_k)
\\ \times
\frac{g(\lambda_\alpha,\lambda_r) g(\lambda_r,\lambda_\beta)}
{f(\lambda_\alpha,\lambda_r)f(\lambda_r,\lambda_\beta)}
f(\lambda_\alpha,\lambda_\beta)
\prod_{\substack{\gamma=1\\ \gamma\ne\alpha,\beta}}^{r}
f(\lambda_\alpha,\lambda_\gamma)f(\lambda_\gamma,\lambda_\beta)
\prod_{\substack{\gamma=1\\ \gamma\ne\alpha,\beta}}^{r}
B(\lambda_\gamma)\ket{\Uparrow}.
\end{multline}
Thus, in this case we have
\begin{multline}
H_N^{(r,r)}
=
Z_N^{-1}\, c^2
\prod_{\alpha=r+1}^{N} a(\lambda_\alpha,\nu_1)\, b(\lambda_\alpha,\nu_N)
\prod_{\alpha=1}^{r-1} b(\lambda_\alpha,\nu_1)\, a(\lambda_\alpha,\nu_N)
\\ \times
\sum_{\substack{\alpha,\beta=1\\ \alpha\ne\beta}}^{r}
\prod_{k=2}^{N-1}a(\lambda_\alpha,\nu_k)
\prod_{k=2}^{N-1}b(\lambda_\beta,\nu_k)
\frac{g(\lambda_\alpha,\lambda_r) g(\lambda_r,\lambda_\beta)}
{f(\lambda_\alpha,\lambda_r)f(\lambda_r,\lambda_\beta)}
f(\lambda_\alpha,\lambda_\beta)
\\ \times
\prod_{\substack{\gamma=1\\ \gamma\ne\alpha,\beta}}^{r}
f(\lambda_\alpha,\lambda_\gamma)f(\lambda_\gamma,\lambda_\beta)
Z_{N-2}\Big(\{\lambda_\delta\}_{\delta=1,\delta\ne\alpha,\beta}^N;
\{\nu_k\}_{k=2}^{N-1}\Big).
\end{multline}
This formula shows that expression \eqref{HNrrZ} is, in fact,
valid for all values of $r_1$ and $r_2$, $1\leq r_1,r_2\leq N$ with no
further restriction.

Substituting now the determinant representation \eqref{ZN}
for the partition
function $Z_N$ and that for the `reduced' partition function $Z_{N-2}$
into \eqref{HNrrZ}, and cancelling whatever possible, we find
\begin{multline}\label{HTT}
H_N^{(r_1,r_2)}=
\frac{c^2 d(\nu_1,\nu_N)\prod\limits_{k=2}^{N-1}
d(\nu_1,\nu_k)d(\nu_k,\nu_N)}
{\prod\limits_{\alpha=1}^{r_1} a(\lambda_\alpha,\nu_1)
\prod\limits_{\alpha=r_1}^{N}b(\lambda_\alpha,\nu_1)
\prod\limits_{\alpha=1}^{r_2} b(\lambda_\alpha,\nu_N)
\prod\limits_{\alpha=r_2}^{N}a(\lambda_\alpha,\nu_N)
\; {\det}_N T}
\\  \times
\sum_{\alpha=1}^{r_1}
\sum_{\substack{\beta=1\\ \beta\ne\alpha}}^{r_2}
(-1)^{N+\alpha+\beta} \epsilon_{\alpha\beta}
\frac{w_{r_1}(\lambda_\alpha)
\tilde w_{r_2}(\lambda_\beta)}{e(\lambda_\beta,\lambda_\alpha)}
\;
{\det}_{N-2} T_{(\alpha,\beta;1,N)}.
\end{multline}
Here $T_{(\alpha,\beta;1,N)}$ denotes the $(N-2)\times(N-2)$
matrix obtained from $T$ by eliminating the $\alpha$-th and
$\beta$-th rows and the first and the last columns
(i.e., its determinant is just the corresponding minor of $T$).
The symbol $\epsilon_{\alpha\beta}$ stands for the sign function
\begin{equation}
\epsilon_{\alpha\beta}=
\begin{cases}
1 &\text{if}\ \alpha>\beta \\
0 & \text{if}\ \alpha=\beta \\
-1 &\text{if}\ \alpha<\beta
\end{cases}.
\end{equation}
The functions $w_r(\lambda)$ and $\tilde w(\lambda)$
are defined as (compare with \eqref{vr})
\begin{align}
w_r(\lambda) &=
\frac{\prod_{\alpha=r+1}^{N} d(\lambda_\alpha,\lambda)
\prod_{\alpha=1}^{r-1} e(\lambda_\alpha,\lambda)
}{\prod_{k=2}^{N-1}b(\lambda,\nu_k)},\
\\
\tilde w_r(\lambda) &=
\frac{\prod_{\alpha=r+1}^{N} d(\lambda,\lambda_\alpha)
\prod_{\alpha=1}^{r-1} e(\lambda,\lambda_\alpha)
}{\prod_{k=2}^{N-1}a(\lambda,\nu_k)}.
\end{align}
Note, that both $w_r(\lambda_\alpha)$ and
$\tilde w_r(\lambda_\alpha)$
vanish if $r<\alpha$; thus the double sum in \eqref{HTT}
can be formally extended to $N$.

Consider now how the homogeneous limit can be performed
in the just obtained expression for the two-point correlation function.
Here it should be mentioned that, contrary to the formulae
for the partition function and one-point correlators, the two-point
boundary correlation function cannot be expressed as a
determinant in the inhomogeneous model. Nevertheless, as we shall show
now this is just a minor inconvenience which can be very efficiently
solved without much efforts.

First of  all we mention that the problem we are actually
facing is that of finding the homogeneous limit for the
following double sum
\begin{equation}
W=\sum_{\alpha,\beta=1}^{N}
(-1)^{N+\alpha+\beta} \epsilon_{\alpha\beta}
F(\lambda_\alpha,\lambda_\beta)\;
{\det}_{N-2}^{} T_{(\alpha,\beta;1,N)}
\end{equation}
where $F(\lambda,\lambda')$ is some function.
Consider the substitution
\begin{equation}
\lambda_\alpha=\lambda+z_\alpha,\qquad \alpha=1,\dots,N.
\end{equation}
In what follows we shall send all $z_\alpha$ (and $\nu_k$) to zero
but before performing the limit we note that
the double sum above can be rewritten, using
\begin{equation}
F(\lambda+z_1,\lambda+z_2)
 =
\Big[\exp(z_\alpha\partial_{\eps_1}) \exp(z_\beta\partial_{\eps_2})
F(\lambda+\eps_1,\lambda+\eps_2)\Big]
\Big|_{\substack{\eps_1=0 \\ \eps_2=0}}\;,
\end{equation}
in a quite formal way as follows
\begin{equation}
W=\begin{vmatrix}
\exp(z_1\partial_{\eps_1})&
t(\lambda+z_1,\nu_2) &\dots & t(\lambda+z_1,\nu_{N-1}) &
\exp(z_1\partial_{\eps_2})\\
\exp(z_2\partial_{\eps_1})&
t(\lambda+z_2,\nu_2) &\dots & t(\lambda+z_2,\nu_{N-1}) &
\exp(z_2\partial_{\eps_2})\\
\hdotsfor{5}\\
\exp(z_N\partial_{\eps_1})&
t(\lambda+z_N,\nu_2) &\dots & t(\lambda+z_N,\nu_{N-1}) &
\exp(z_N\partial_{\eps_2})
\end{vmatrix}
F(\lambda+\eps_1,\lambda+\eps_2)
\Big|_{\substack{\eps_1=0 \\ \eps_2=0}}
\end{equation}
so that the homogenous limit of expression \eqref{HTT}
can be performed exactly along
the lines given in \cite{ICK-92}.

Namely, we shall put $\nu_k$'s and $z_\alpha$'s to zero in the order
$\nu_1,\ldots,\nu_N,z_1,\ldots,z_N$ each time keeping the leading
order. The prefactor in \eqref{HTT} becomes
\begin{multline}
\frac{\sin^2(2\eta)}{
[\sin(\lambda+\eta)]^{N+r_1-r_2+1}[\sin(\lambda-\eta)]^{N+r_2-r_1+1}\,
{\det}_N^{}\varPhi}
\\ \times
\frac{(N-1)!\, (N-2)!}
{\Big(z_2\cdot \frac{z_3^2}{2!}\cdot\frac{z_4^3}{3!}
\cdots\frac{z_N^{N-1}}{(N-1)!} \Big)
\Big(\nu_3\cdot \frac{\nu_4^2}{2!}\cdots
\frac{\nu_{N-1}^{N-2}}{(N-3)!} \Big)}
\end{multline}
while the double sum goes into
\begin{multline}
\bigg(z_2\cdot \frac{z_3^2}{2!}\cdot\frac{z_4^3}{3!}
\cdots\frac{z_N^{N-1}}{(N-1)!} \bigg)
\bigg(\nu_3\cdot \frac{\nu_4^2}{2!}\cdots
\frac{\nu_{N-1}^{N-2}}{(N-3)!} \bigg)
\\  \times
{\det}\Big(\varPhi_{\alpha,k}\Big|
\partial_{\eps_2}^{\alpha-1}\Big|\partial_{\eps_1}^{\alpha-1}
\Big)_{1\leq \alpha \leq N,1\leq k\leq N-2}\;
h_N^{(r_1,r_2)}(\eps_1,\eps_2)\bigg|_{\eps_1=\eps_2=0}
\end{multline}
where
\begin{equation}\label{h2}
h_N^{(r_1,r_2)}(\eps_1,\eps_2)=
\frac{(\sin\eps_1)^{N-r_1}[\sin(\eps_1-2\eta)]^{r_1-1}
(\sin\eps_2)^{N-r_2}[\sin(\eps_2+2\eta)]^{r_2-1}}
{\sin(\eps_2-\eps_1+2\eta)[\sin(\eps_1+\lambda-\eta)]^{N-2}
[\sin(\eps_2+\lambda+\eta)]^{N-2}}.
\end{equation}
Thus, in the homogeneous limit for the two-point correlation function
we obtain
\begin{multline}\label{H2hom}
H_N^{(r_1,r_2)}=
\frac{(N-1)!\,(N-2)!\,\sin^2(2\eta)}
{\big[\sin(\lambda+\eta)\big]^{N+r_1-r_2+1}
\big[\sin(\lambda-\eta)\big]^{N+r_2-r_1+1}
{\det}_N \varPhi}
\\  \times
\left[
{\det}\Big(\varPhi_{\alpha,k}\Big|
\partial_{\eps_2}^{\alpha-1}\Big|\partial_{\eps_1}^{\alpha-1}
\Big)_{1\leq \alpha \leq N,1\leq k\leq N-2}\;
h_N^{(r_1,r_2)}(\eps_1,\eps_2)\right]\bigg|_{\eps_1=\eps_2=0}.
\end{multline}
This determinant representation is analogous
to those of the previous section for one-point
correlation functions.

\section{Orthogonal polynomials representation}\label{sec-rslt}

In this section the results for one- and two-point
boundary correlation functions
will be analyzed by making
use of the orthogonal polynomials theory, along the lines proposed
in paper \cite{CP-05}. Here we show that the two-point boundary
correlation function, studied in the previous section,
is expressible in terms of one-point ones.

Let $\{P_n(x)\}_{n=0}^\infty$ be a set of polynomials, with
non-vanishing leading coefficient
\begin{equation}
P_n(x)=\kappa_n x^n + \dots,\qquad kappa_n\ne0,
\end{equation}
and orthogonal on the real axis with respect to some
weight $\mu(x)$,
\begin{equation}\label{ortho}
\int_{-\infty}^{\infty} P_{n_1}(x)\, P_{n_2}(x)\, \mu(x)\, \mathrm{d} x
=h_{n_1} \delta_{n_1n_2}.
\end{equation}
Let $c_n$ denote $n$-th  moment of the weight $\mu(x)$, i.e.
\begin{equation}
c_n =\int_{-\infty}^{\infty} x^n \mu(x) \mathrm{d} x,\qquad
n=0,1,\ldots
\end{equation}
and let us consider the $(n+1)\times (n+1)$ determinant
\begin{equation}
\varDelta_n =
\begin{vmatrix}
c_0&c_1& \dots &c_n \\
c_1& c_2&  \dots & c_{n+1}\\
\hdotsfor{4}  \\
c_n & c_{n+1} & \dots & c_{2n}
\end{vmatrix}.
\end{equation}
Using the orthogonality condition \eqref{ortho} and
well-known properties of determinants, one can easily
find the following formula
\begin{equation}\label{hankel}
\varDelta_n=\prod_{k=0}^n \frac{h_k}{\kappa_k^2}.
\end{equation}
This formula can be used for computation of determinants, provided
the orthogonal polynomials $\{P_n(x)\}_{n=0}^\infty$ are known.
On the other hand, the polynomials $\{P_n(x)\}_{n=0}^\infty$
can in turn be expressed
as determinants. For later use let us introduce the notation
\begin{equation}
\varDelta_n^{(k)}(x_1,\dots,x_k) =
\begin{vmatrix}
c_0&c_1& \dots &c_{n-k}  & 1 & 1 & \dots & 1 \\
c_1&c_2& \dots &c_{n-k+1}& x_1 & x_2 &\dots &x_k \\
\hdotsfor{8}  \\
c_n& c_{n+1}&\dots & c_{2n-k} & x_1^n & x_2^n &\dots &x_k^n
\end{vmatrix}
\end{equation}
so that $\varDelta_n^{(0)}\equiv\varDelta_n$.
For the polynomials one can find that
\begin{equation}\label{Px}
P_n(x)=\frac{\kappa_n}{\varDelta_{n-1}}\varDelta_n^{(1)}(x).
\end{equation}
For a proof, see, e.g., book \cite{S-75}.

The relation \eqref{Px} can be read off inversely thus giving
an expression for the determinant $\varDelta_n^{(1)}(x)$
in terms of the polynomials $P_n(x)$. Taking into account that
(see \eqref{hankel})
\begin{equation}\label{kh}
\frac{h_n}{\kappa_n^2}= \frac{\varDelta_n}{\varDelta_{n-1}}
\end{equation}
we can write
\begin{equation}\label{D1}
\frac{\varDelta_n^{(1)}(x)}{\varDelta_n}= \frac{\kappa_n}{h_n}\; P_n(x).
\end{equation}

Consider now the case of $\varDelta_n^{(2)}(x_1,x_2)$.
It is clear that the term of the highest powers on both
$x_1$ and $x_2$ is just
$\varDelta_{n-2}(x_2^{n}x_1^{n-1}-x_1^{n} x_2^{n-1})$; extending further
the  methods of \cite{S-75} we obtain
\begin{equation}
\varDelta_n^{(2)}(x_1,x_2)=
\frac{\varDelta_{n-2}}{\kappa_n\kappa_{n-1}}
\big[P_{n-1}(x_1)P_n(x_2)-P_{n}(x_1)P_{n-1}(x_2)\big].
\end{equation}
Again using \eqref{kh}, we write
\begin{align}\label{D2}
\frac{\varDelta_n^{(2)}(x_1,x_2)}
{\varDelta_n}
&= \frac{\kappa_n\kappa_{n-1}}{h_nh_{n-1}}\;
\big[P_{n-1}(x_1)P_n(x_2)-P_{n}(x_1)P_{n-1}(x_2)\big]
\notag\\
& = \frac{\kappa_n\kappa_{n-1}}{h_nh_{n-1}}\;
\begin{vmatrix}
P_{n-1}(x_1) & P_{n}(x_1) \\
P_{n-1}(x_2) & P_{n}(x_2)
\end{vmatrix}.
\end{align}
This formula can be easily extended to the general case of
$\varDelta^{(n)}(x_1,\dots,x_n)$; in what follows we shall make use
only of formulae \eqref{D1} and \eqref{D2}.

Consider now how we can use all these formulae in application
to the boundary correlation functions. First we note, following
paper \cite{CP-05}, that the determinant entering the expression for the
homogenous model partition function can be related with orthogonal
polynomials using the integral representation
\begin{equation}\label{int}
\frac{\sin(2\eta)}{\sin(\lambda-\eta)\sin(\lambda+\eta)} =
\int_{-\infty}^{\infty} \rme^{x(\lambda-\pi/2)}
\frac{\sinh(\eta x)}{\sinh(\pi x/2)}\; \mathrm{d}x.
\end{equation}
This formula is valid if $0<\eta<\pi/2$ and $\eta<\lambda<\pi-\eta$;
these values of $\lambda$ and $\eta$ correspond
to the so-called disordered regime of the six-vertex model
(for similar formulae valid for other regimes, see \cite{Zj-00}).
This regime is the most interesting especially
for combinatorial applications of the six-vertex model with DWBC
(in these applications one further specializes to $\lambda=\pi/2$).
It can be easily seen that our results below, however,
do not depend on the particular choice of the regime, and
can be extended to other regimes simply using the proper analytical
continuation in the parameters $\lambda$ and $\eta$.

Formula \eqref{int} implies that we have to deal with
the set of polynomials which are orthogonal with respect to
the following weight function
\begin{equation}
\mu(x)=\mu(x;\lambda,\eta)=
\rme^{x(\lambda-\pi/2)}
\frac{\sinh(\eta x)}{\sinh(\pi x/2)}.
\end{equation}
The corresponding polynomials $P_n(x)=P_n(x;\lambda,\eta)$
also depend on $\lambda$ and $\eta$ which are to be considered
as parameters. For later use let us mention the following
useful property of these polynomials
\begin{equation}\label{cross}
P_n(x;\lambda,\eta)=(-1)^n\, P_n(-x;\pi-\lambda,\eta).
\end{equation}
This property can be easily established in virtue of formula
\eqref{Px}. It is to mentioned also that both the leading coefficient
$\kappa_n=\kappa_n(\lambda,\eta)$ and the normalization constant
$h_n=h_n(\lambda,\eta)$ are invariant under the substitution
$\lambda\to\pi-\lambda$.

The transformation $\lambda\to\pi-\lambda$ is related to the so-called
crossing symmetry of the six-vertex model
which has useful consequences for the one-point
boundary correlation function $H_N^{(r)}$. Recall that the
the crossing symmetry is the symmetry of the vertex weights
under reflection with respect to the vertical axis, and simultaneous
interchange of the functions $a$ and $b$, which is equivalent
to setting $\lambda\to\pi-\lambda$.
As we shall explain now, these simple
properties related to the crossing symmetry
allow one to derive easily two equivalent representations for
the one-point boundary correlation function $H_N^{(r)}$. Using
these formulae we shall then show that
the two-point boundary correlation function
$H_N^{(r_1,r_2)}$ is expressible in terms of  one-point
boundary correlators.

Indeed, since the lattice with DWBC is
invariant under the reflection with respect to
the vertical axis, the crossing symmetry thus imply that the
following relation holds
\begin{equation}\label{crossym}
H_N^{(r)}(\lambda,\eta)=H_N^{(N-r+1)}(\pi-\lambda,\eta).
\end{equation}
Consider expression \eqref{Hhom} for the one-point function
$H_N^{(r)}$. Due to \eqref{D1} we can rewrite it as
\begin{multline}\label{HNlong}
H_N^{(r)}(\lambda,\eta)=\frac{(N-1)!\,\sin(2\eta)}
{\big[\sin(\lambda+\eta)\big]^r\big[\sin(\lambda-\eta)\big]^{N-r+1}}\;
\frac{\kappa_{N-1}(\lambda,\eta)}{h_{N-1}(\lambda,\eta)}
\\ \times
P_{N-1}(\partial_\eps;\lambda,\eta)
\frac{(\sin\eps)^{N-r}[\sin(\eps-2\eta)]^{r-1}}
{[\sin(\eps+\lambda-\eta)]^{N-1}}\bigg|_{\eps=0}.
\end{multline}
Taking into account \eqref{cross} and the properties of the
leading coefficient $\kappa_n(\lambda,\eta)$
and the normalization constant $h_n(\lambda,\eta)$ mentioned above,
it can be easily seen that from \eqref{crossym} and \eqref{HNlong}
the following expression is valid as well
\begin{multline}\label{HNlong2}
H_N^{(r)}(\lambda,\eta)=\frac{(N-1)!\,\sin(2\eta)}
{\big[\sin(\lambda+\eta)\big]^r\big[\sin(\lambda-\eta)\big]^{N-r+1}}\;
\frac{\kappa_{N-1}(\lambda,\eta)}{h_{N-1}(\lambda,\eta)}
\\ \times
P_{N-1}(\partial_\eps;\lambda,\eta)
\frac{(\sin\eps)^{r-1}[\sin(\eps+2\eta)]^{N-r}}
{[\sin(\eps+\lambda+\eta)]^{N-1}}\bigg|_{\eps=0}.
\end{multline}
Note that this expression means simply that the limit $\eps\to 0$ in
\eqref{HNlong} can be changed into $\eps\to 2\eta$ without altering
the result.

Thus, these two equivalent representations,
\eqref{HNlong} and \eqref{HNlong2}, can be used in the study of
the two-point
correlation function $H_N^{(r_1,r_2)}$ given by
expression  \eqref{H2hom}, which certainly involves
similar structures.

Before turning to this analysis, let us put the above formulae
for the one-point function in a more compact and convenient
notations. In what follows we shall often omit the dependence
on $\lambda$ and $\eta$ where possible.

We define the functions
\begin{equation}
\omega(\epsilon)=\frac{\sin(\lambda+\eta)}{\sin(\lambda-\eta)}\,
\frac{\sin\eps}{\sin(\eps-2\eta)},\qquad
\varrho(\epsilon)=\frac{\sin(\lambda-\eta)}{\sin(2\eta)}\,
\frac{\sin(\eps-2\eta)}{\sin(\eps+\lambda-\eta)};
\end{equation}
which are related to each other as
\begin{equation}\label{go}
\varrho(\eps)=\frac{1}{\omega(\eps)-1}.
\end{equation}
Also we define
\begin{equation}
\tilde\omega(\epsilon)=\frac{\sin(\lambda-\eta)}{\sin(\lambda+\eta)}\,
\frac{\sin\eps}{\sin(\eps+2\eta)},\qquad
\tilde\varrho(\epsilon)=\frac{\sin(\lambda+\eta)}{\sin(2\eta)}\,
\frac{\sin(\eps+2\eta)}{\sin(\eps+\lambda+\eta)};
\end{equation}
which are in turn related to each other as
\begin{equation}\label{tgto}
\tilde\varrho(\eps)=\frac{1}{1-\tilde\omega(\eps)}.
\end{equation}
Note, that the functions with tildes are introduced such that
\begin{equation}
\tilde\omega(\eps;\lambda,\eta)=\omega(-\eps;\pi-\lambda,\eta),\qquad
\tilde\varrho(\eps;\lambda,\eta)=-\varrho(-\eps;\pi-\lambda,\eta)
\end{equation}
in accordance with the crossing symmetry considerations made above.
Additionally, let us denote
\begin{equation}
K_{N-1}(x)=(N-1)!\,\varphi^N\, \frac{\kappa_{N-1}}{h_{N-1}}\;
P_{N-1}(x)
\end{equation}
where $\varphi=\varphi(\lambda,\eta)$ is exactly the function
defining entries of the matrix $\varPhi$, see \eqref{varphi}.
In these notations formulae \eqref{HNlong} and \eqref{HNlong2}
for the correlation function $H_N^{(r)}$ read
\begin{equation}
H_N^{(r)}=K_{N-1}(\partial_\eps)
\big[\omega(\eps)\big]^{N-r}[\varrho(\eps)]^{N-1}\Big|_{\eps=0}
\end{equation}
and
\begin{equation}
H_N^{(r)}
= K_{N-1}(\partial_\eps)
\big[\tilde\omega(\eps)\big]^{r-1}
[\tilde\varrho(\eps)]^{N-1}\Big|_{\eps=0},
\end{equation}
respectively.

Consider now our main object, the two-point correlation function
$H_N^{(r_1,r_2)}$, which is given by  formula \eqref{H2hom}.
Obviously, function \eqref{h2} contains all the
structures introduced above apart from the factor
$\sin(\eps_2-\eps_1+2\eta)$ standing in the denominator there.
However, using the identity
\begin{equation}
\sin(2\eta) \sin(\eps_2-\eps_1+2\eta)=
\sin\eps_1 \sin\eps_2-
\sin(\eps_1-2\eta)\sin(\eps_2+2\eta)
\end{equation}
it can be easily seen that
\begin{equation}
\frac{\sin(\eps_1+\lambda-\eta)
\sin(\eps_2+\lambda+\eta)}{\sin(\eps_2-\eps_1+2\eta)}
=\frac{1}{\varphi \varrho(\eps_1) \tilde \varrho(\eps_2)}\;
\frac{1}{\omega(\eps_1)\tilde\omega(\eps_2)-1}.
\end{equation}
Thus, taking into account formula \eqref{D2}
we can write the two-point correlation function in the form
\begin{multline}\label{H2nice}
H_N^{(r_1,r_2)}=
\left[K_{N-1}(\partial_{\eps_1})K_{N-2}(\partial_{\eps_2})
-K_{N-2}(\partial_{\eps_1})K_{N-1}(\partial_{\eps_2})\right]
\\ \times
\frac{\big[\omega(\eps_1)\big]^{N-r_1}[\varrho(\eps_1)]^{N-2}
\big[\tilde\omega(\eps_2)\big]^{N-r_2}[\tilde\varrho(\eps_2)]^{N-2}}
{\omega(\eps_1)\tilde\omega(\eps_2)-1}\bigg|_{\eps_1=0,\eps_2=0}.
\end{multline}
Now taking into account that $\omega(\eps),\tilde \omega(\eps)\to 0$
as $\eps\to 0$
we can expand the denominator in \eqref{H2nice} in power series
and it can be easily seen that only the first few terms
(actually not more than $N$) of this expansion will contribute.
As a result, in virtue
of relations \eqref{go} and \eqref{tgto}, we arrive
to the following expression in terms of the one-point functions
\begin{multline}\label{H=HH}
H_N^{(r_1,r_2)}=
\sum_{j=1}^{N}
\Big(
H_N^{(r_1-j+1)} H_{N-1}^{(N-r_2+j)}
-H_N^{(r_1-j)} H_{N-1}^{(N-r_2+j)}
\\
-H_{N-1}^{(r_1-j)} H_{N}^{(N-r_2+j+1)}
+H_{N-1}^{(r_1-j)} H_{N}^{(N-r_2+j)}
\Big)
\end{multline}
where it is assumed that if $r\leq0$ or $r\geq N+1$ then
$H_N^{(r)}=0$ by definition. The formula \eqref{H=HH}
is our main result here.

As a comment to this result let us
rewrite it in terms of the generating functions.
Let us introduce the generating function for the two-point
correlation function
\begin{equation}
H_N(u,v):= \sum_{r=1,s=1}^{N} H_N^{(N-r+1,s)} u^{r-1} v^{s-1}
\end{equation}
and define the generating function for the one-point correlation
function
\begin{equation}
H_N(u):= \sum_{r=1}^{N} H_N^{(N-r+1)} u^{r-1}.
\end{equation}
Then \eqref{H=HH} implies that
\begin{align}\label{HNuv}
H_N(u,v)&=\frac{(u-1)H_N(u)\cdot vH_{N-1}(v) -
uH_{N-1}(u)\cdot (v-1)H_N(v)}{u-v}
\notag\\
&=\frac{1}{u-v}
\begin{vmatrix}
(u-1)H_{N}(u) & uH_{N-1}(u)\\
(v-1)H_{N}(v) & vH_{N-1}(v)
\end{vmatrix}.
\end{align}
This formula generalize to arbitrary values of the vertex weights
the result of paper \cite{S-02}
where an  equivalent expression was derived
in the case  $\lambda=\pi/2$ and $\eta=\pi/6$,
i.e., when $a=b=c$
(the so-called ice point).

As a final comment we would like to stress that formula \eqref{H=HH}
implies that doubly refined weighted enumerations of ASMs can be easily
obtained from the corresponding singly refined ones.
In particular, the explicit expressions for the
refined $x$-enumerations, known for $x=1,2,3$,
can be just plugged into this formula to obtain
the corresponding doubly refined ones.
A discussion of the related one-point boundary correlation functions
in application to the refined $1$-, $2$-, and
$3$-enumerations of ASMs can be found in \cite{CP-05}.

\section*{Acknowledgments}

We acknowledge financial support from MIUR PRIN programme (SINTESI
2004). One of us (AGP) is also supported in part by Civilian Research
and Development Foundation (under CRDF grant RUM1-2622-ST-04), by
Russian Foundation for Basic Research (under RFFI grant
04-01-00825), and by the programme Mathematical Methods in Nonlinear
Dynamics of Russian Academy of Sciences. This work is partially
done within the European Community network EUCLID (HPRN-CT-2002-00325).



\begin{thebibliography}{**}
\bibitem{S-41}
J.C. Slater.
Theory of the transition in KH${}_2$PO${}_4$.
\textit{Journ. Chem. Phys.} \textbf{9} (1941) 16--33.

\bibitem{L-67}
E.H. Lieb.
Exact solution of the F model of an antiferroelectric.
\textit{Phys. Rev. Lett.} \textbf{18} (1967) 1046--1048.

\bibitem{L-67a}
E.H. Lieb.
Exact solution of the two-dimensional Slater
KDP model of a ferroelectric.
\textit{Phys. Rev. Lett.} \textbf{19} (1967) 108--110.

\bibitem{S-67}
B. Sutherland.
Exact solution of a two-dimensional model for hydrogen-bonded crystals.
\textit{Phys. Rev. Lett.} \textbf{19} (1967) 103--104.

\bibitem{LW-72}
E.H. Lieb and F.Y. Wu.
In: \textit{Phase Transitions and Critical Phenomena}, Vol. 1,
edited by C. Domb and M. S. Green, Academic Press, London, 1972,
p.~321.

\bibitem{B-82}
R.J. Baxter.
\textit{Exactly Solved Models in Statistical Mechanics}.
Academic press, San Diego, 1982.

\bibitem{K-82}
V.E. Korepin.
Calculations of norms of Bethe wave functions.
\textit{Commun. Math. Phys.} (1982) \textbf{86} 391--418.

\bibitem{KBI-93}
V.E. Korepin, N.M. Bogoliubov, and A.G. Izergin.
\textit{Quantum Inverse Scattering Method and Correlation Functions}.
Cambridge University Press, Cambridge, 1993.

\bibitem{I-87}
A.G. Izergin. Partition function of the six-vertex model in
the finite volume.
\textit{Sov. Phys. Dokl.} \textbf{32} (1987) 878--879.

\bibitem{ICK-92}
A.G. Izergin, D.A. Coker, and V.E. Korepin.
Determinant formula for the six-vertex model.
\textit{J. Phys. A: Math. Gen.} \textbf{25} (1992) 4315--4334.

\bibitem{MRR-83}
W.H. Mills, D.P. Robbins, and H. Rumsey.
Alternating-sign matrices and descending plane partitions.
\textit{J. Combin. Theory Ser. A} \textbf{34} (1983) 340--359.

\bibitem{EKLP-92a}
N. Elkies, G. Kuperberg, M. Larsen, and J. Propp.
Alternating-sign matrices and domino tilings (part I).
\textit{J. Algebraic Combin.} {\bf 1} (1992) 111--132.

\bibitem{EKLP-92b}
N. Elkies, G. Kuperberg, M. Larsen, and J. Propp.
Alternating-sign matrices and domino tilings (part II).
\textit{J. Algebraic Combin.} {\bf 1} (1992) 219--234.

\bibitem{Z-96a}
D. Zeilberger.
Proof of the alternating sign matrix conjecture.
\textit{Elec. J. Comb.} \textbf{3} (2) (1996) R13.

\bibitem{Ku-96}
G. Kuperberg.
Another proof of the alternative-sign matrix conjecture.
\textit{Internat. Math. Res. Notices} \textbf{1996} (1996) 139--150.


\bibitem{Z-96b}
D. Zeilberger.
Proof of the refined alternating sign matrix conjecture.
\textit{New York J. Math} \textbf{2} (1996) 59--68.

\bibitem{Br-99}
D.M. Bressoud.
\textit{Proofs and Confirmations: The Story of the
Alternating Sign Matrix Conjecture}.
Cambridge University Press, Cambridge, 1999.


\bibitem{RS-01}
A.V. Razumov and  Yu.G. Stroganov.
Spin chains and combinatorics.
\textit{J. Phys. A: Math. Gen.} \textbf{34} (2001) 3185--3190.

\bibitem{dGR-04}
J. de Gier and V. Rittenberg.
Refined Razumov-Stroganov conjectures for open boundaries.
\textit{J. Stat. Mech: Theor. Exp.} (2004) P09009.

\bibitem{DfZj-04}
P. Di Francesco and P. Zinn-Justin.
Around the Razumov-Stroganov conjecture: proof of a multi-parameter
sum rule.
\textit{Elec. J. Comb.} \textbf{12} (1) (2005) R6.

\bibitem{dGN-05}
J. de Gier and B. Nienhuis.
Brauer loops and the commuting variety.
\textit{J. Stat. Mech: Theor. Exp.} (2005) P01006.

\bibitem{BKZ-02}
N.M. Bogoliubov, A.V. Kitaev, and M.B. Zvonarev.
Boundary polarization in the six-vertex model.
\textit{Phys. Rev. E} \textbf{65} (2002) 026126.

\bibitem{BPZ-02}
N.M. Bogoliubov, A.G. Pronko, and M.B. Zvonarev.
Boundary correlation functions of the six-vertex model.
\textit{J. Phys. A: Math. Gen.} \textbf{35} (2002) 5525--5541.

\bibitem{FP-04}
O. Foda and I. Preston.
On the correlation functions of the domain wall six-vertex model.
\textit{J. Stat. Mech: Theor. Exp.} (2004) P11001.

\bibitem{S-02}
Yu. G. Stroganov.
A new way to deal with Izergin--Korepin determinant at root of unity.
Preprint, arXiv: math-ph/0204042.

\bibitem{Df-04}
P. Di Francesco. A refined Razumov-Stroganov conjecture II.
\textit{J. Stat. Mech: Theor. Exp.} (2004) P11004.

\bibitem{CP-05}
F. Colomo and A.G. Pronko.
Square ice, alternating sign matrices,
and classical orthogonal polynomials.
\textit{J. Stat. Mech: Theor. Exp.} (2005) P01005.

\bibitem{Zj-00}
P. Zinn-Justin.
Six-vertex model with domain wall boundary conditions
and one-matrix model.
\textit{Phys. Rev. E} \textbf{62} (2000) 3411--3418.

\bibitem{S-75}
G. Szeg\"o.
\textit{Orthogonal Polynomials}.
Fourth edition, Colloquium Publications,
Vol. 23, Amer. Math. Soc., Providence, RI, 1975.

\end{thebibliography}
\end{document}